\documentclass{osa-article}
\usepackage{graphicx}
\usepackage{booktabs}
\journal{oe}

\articletype{Research Article}

\begin{document}

\title{Hadamard `Pipeline' Coding Computational Ghost Imaging}

\author{Cheng Zhou,\authormark{1,2,3} Xin Yu,\authormark{1} Xiwei Zhao,\authormark{3}  Heyan Huang,\authormark{4} Gangcheng Wang,\authormark{1} Xue Wang,\authormark{3} Lijun Song,\authormark{2,5} and Kang Xue\authormark{1,6}}

\address{\authormark{1}Center for Quantum Sciences and School of Physics, Northeast Normal University, Changchun 130024, China\\
\authormark{2}Institute for Interdisciplinary Quantum Information Technology, Jilin Engineering Normal University, Changchun 130052, China\\
\authormark{3}Jilin Engineering Laboratory for Quantum Information Technology, Changchun 130052, China\\
\authormark{4}College of Science, Shanghai Institute of Technology, shanghai 201418, China}

\email{\authormark{5}ccdxslj@126.com} 
\email{\authormark{6}xuekang@nenu.edu.cn} 


\begin{abstract}
The Hadamard matrix with orthogonality is a more important modulation matrix for computational ghost imaging (CGI), especially its optimized Hadamard matrix. However, as far as we know, little mention has been paid to efficient and convenient Hadamard matrix generation for CGI. The existing methods are to reconstruct any row of Hadamard matrix into two-dimensional matrix and then optimize it. In this work, we propose a Hadamard `pipeline' coding computational ghost imaging approach, which can directly generate two-dimensional Hadamard derived pattern and Hadamard optimization sequence, whereby both the memory consumption and the complexity of coding implementation for CGI can be significantly reduced. The optimization method of commonly used hadamard optimization sequence implementation is also discussed. This method provides a new approach for Hadamard sequence optimization and ghost imaging applications.
\end{abstract}

\section{Introduction}
Ghost imaging (GI) is a novel imaging technique, which is realized by correlating the total light signal reflected (or transmitted) from the target collected by the single-pixel detector with the reference spatial light distribution signal collected by the multi-pixel array detector. In 2008, computational ghost imaging (CGI) theoretical scheme that requires only one single-pixel detector was proposed by Shapiro \cite{tcgi} and later was experimental demonstrated by Bromberg \cite{ecgi}. The CGI scheme is closely related to single-pixel imaging, which requires spatial light coding, light field modulation equipment and single-pixel detector \cite{sun2017russian}. Among them, spatial light coding is the key factor affecting the CGI imaging efficiency. Also, various types of spatial light coding for CGI have been proposed and used, such as random speckle coding (binary or gray level), Hadamard coding \cite{sun2017russian}, sinusoidal coding \cite{MahdiKhamoushi:15,zhang2015single,zhang2017fast}, etc.

Recently, CGI based on Hadamard coding has attracted more and more researchers' attention, and has aroused in creasing interest in application like microscopic imaging \cite{gt,gc}, X-ray imaging \cite{ftxgi,exgi,lsxgi,txgi,xmi1}, three-dimensional imaging \cite{sun20133d,tdgi,rs2016}, methane gas detection \cite{Gibson:17} and so on \cite{Li:13,Liu:15,khakimov2016ghost,PhysRevLett.121.114801,Zhou:19}. The most important reason is that Hadamard derived patterns are orthogonal to each other, particularly the final reconstruction quality of CGI is significantly improved.

However, the mid results of natural order Hadamard CGI contain overlaps before the whole set of patterns are completed employed. Hence, the optimized ordering of Hadamard coding schemes \cite{sun2017russian,Zhou_2019,yu2019single} were proposed to improve CGI's imaging performance. Typically, Sun et al \cite{sun2017russian} proposed a `Russian dolls' ordering of Hadamard coding scheme, which reorderd the Hadamard matrix from low to high order according to high order including low order characteristics and transpose relationships, whereby the optimal result for any truncation of that pattern sequence can be achieved. Subsequently, a multi-resolution progressive computational ghost imaging (MPCGI) scheme was reported by Zhou et al \cite{Zhou_2019}. The MPCGI scheme can achieve continuous multi-resolution imaging directly and quickly via reordered the Hadamard matrix from low to high by even order. More practically, Yu et al \cite{yu2019single} proposed a hadamard origami pattern construction framework, which the sampling rate can be reduced to 0.5\% by symmetric reverse folding, axial symmetry and partial pattern order adjustment. Nevertheless, we find that it is not the ultimate optimal solution as the complexity of Hadamard coding required to implement the above scheme can be further reduced.

In order to meet the practical application requirements more closely, a low memory, simple and efficient optical field optimization coding method which is more suitable for ghost imaging experiments is urgently needed. To some extent, the optimization of hadamard coding may be an effective solution. Therefore, based on the previous work of Hadamard optimization coding, we propose a new Hadamard `pipeline' coding method for CGI. This method can directly generate two-dimensional Hadamard derived pattern through four fixed transformation rules. At the same time, the complexity of two-dimensional Hadamard derived pattern generation and the memory consumption of the computer are greatly reduced, which is more conducive to the optimization of Hadamard matrix and increases the flexibility of using Hadamard derived patte to realize high-performance CGI. Moreover, according to the Hadamard matrix generation rules, we also demonstrate an optimization method to simply obtain `Russian dolls' ordering and MPCGI ordering.

\section{ CGI based on Hadamard Coding Principle}

\begin{figure}[htbp]
\centering
\includegraphics[width=8cm]{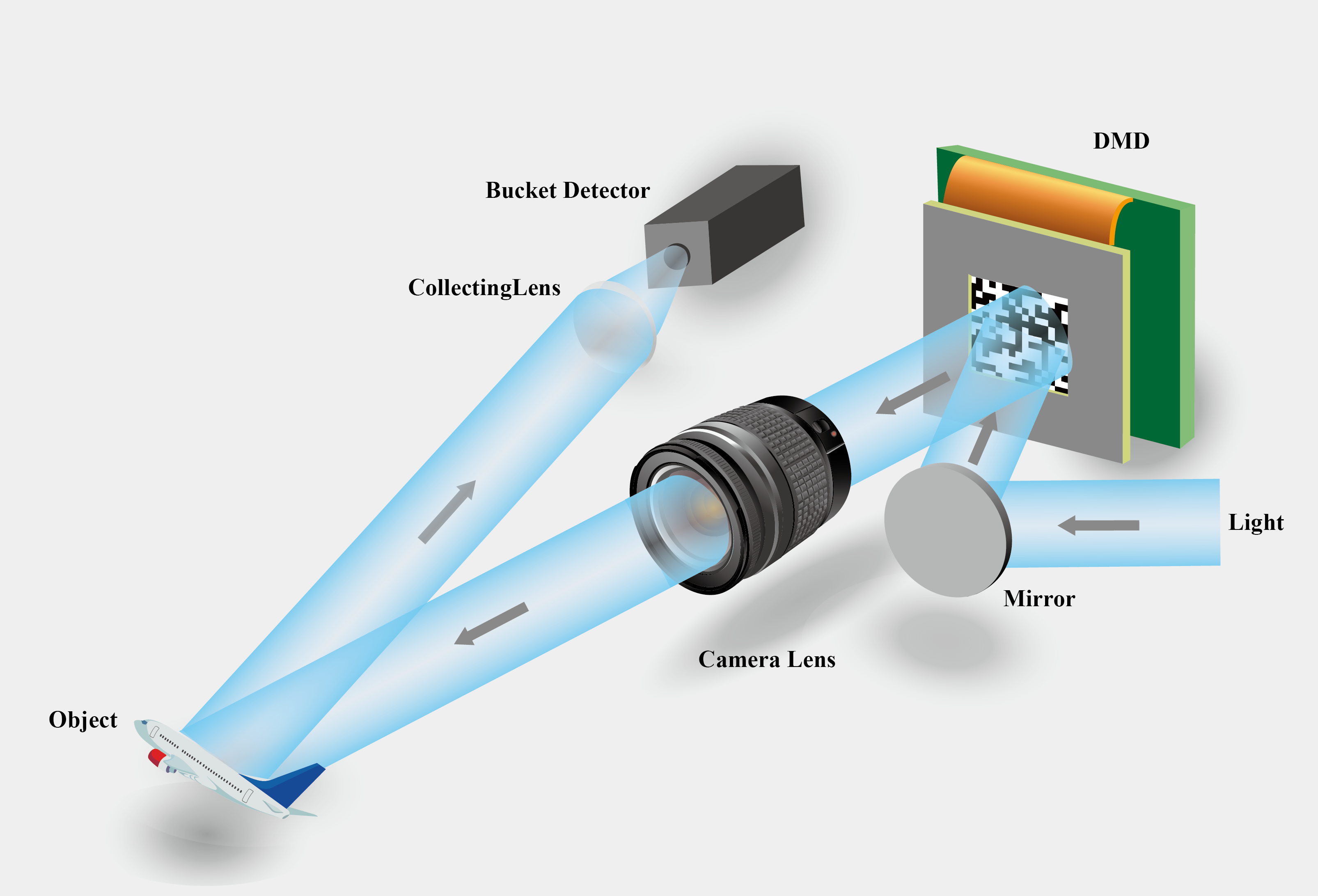}
\caption{The diagram of computational ghost imaging.}
\label{escheme}
\end{figure}

 The CGI system configuration is illustrated in Fig.\ref{escheme}, which includes a camera lens, a digital micromirror devices (DMD), a reflecting mirror, a collecting lens and a bucket detector. The light field $I^{(m)}(x,y)$ modulated by DMD is imaged onto the target object by camera lens.  The bucket detector is placed on one of the reflection orientations to make the measurement of total light signal $B^{(m)}$.

 The target object can be obtained by computing the correlation between $I^{(m)}(x,y)$ and $B^{(m)}$:

\begin{eqnarray}\label{eqtgi}
O_{(x,y)} = \langle B^{(m)}I^{(m)}_{(x,y)}\rangle-\langle B^{(m)}\rangle\langle I^{(m)}_{(x,y)}\rangle, m=1,2,3,\ldots,M,
\end{eqnarray}
where $\langle \cdot \rangle=\frac{1}{M}\sum^{M}_{m=1}(\cdot)$ and $m$ is the number of measurement times. 

\subsection{Traditional two-dimensional hadamard derived coding method (THDC)}

The Hadamard basis is a square matrix composed of $+1$ and $-1$, and can be generated rapidly by Kronecker product, that is:

\begin{equation}
   H_{2^1}=\left[\begin{array}{cc}
                  +1 & +1 \\
                  +1 & -1
           \end{array}\right],
\end{equation}
\begin{equation}
   H_{2^k}=H_{2^1}\otimes H_{2^{k-1}}\\=\left[\begin{array}{cc}
                                               +H_{2^{k-1}} & +H_{2^{k-1}} \\
                                               +H_{2^{k-1}} & -H_{2^{k-1}}
                                      \end{array}\right],\\
\end{equation}
where $2<k$ (integer),  and $ \otimes $ denotes the Kronecker product.
Hence, a Hadamard matrix of size $ M\times N$($M=N$),
\begin{equation}
H_{2^k}(m,n)=\left[\begin{array}{cccc}
               H{(1,1)} & H{(1,2)} & \cdots & H{(1,N)} \\
               H{(2,1)} & H{(2,2)} & \cdots & H{(2,N)} \\
                \vdots& \vdots & \ddots & \vdots \\
               H{(M,1)} & H{(M,2)} & \cdots & H{(M,N)}
             \end{array}\right].
\end{equation}
where, $m=1,2,3,\ldots,M, n=1,2,3,\ldots,N$. To construct the spatial light coding matrix for CGI, the $m$th row in the Hadamard matrix $H_{2^k}(m,n)$ can be reshaped to form a two-dimensional square Hadamard derived pattern, namely:

 \begin{equation}
 H_{2^k}^{(m)}(p,p)=\left[\begin{array}{cccc}
               H{(1,1)} & H{(1,2)} & \cdots & H{(1,p)} \\
               H{(2,1)} & H{(2,2)} & \cdots & H{(2,p)} \\
                \vdots& \vdots & \ddots & \vdots \\
               H{(p,1)} & H{(p,2)} & \cdots & H{(p,p)}
             \end{array}\right].
 \end{equation}
where, $m=1,2,3,\ldots,M, p\times p=N$. These mathematical operations lead to an ordering of the rows, which is called `natural order'.

THDC method is the most common means for optical field coding in CGI, though maybe not the best one. Deserved to be mentioned, although we can first generate a high-order Hadamard matrix and then reshape each row into a two-dimensional matrix for GI experiments,  we need to regenerate a higher-order Hadamard matrix (contains the lower order Hadamard matrix) when the imaging resolution is insufficient (the number of images used for experiments is insufficient). Thus this is wasteful, why don't we directly generate two-dimensional Hadamard derived matrix that can be used for experiments? If the number of images is not enough, why not generate new Hadamard derived pattern directly?

\subsection{New two-dimensional Hadamard `pipeline' coding method (NHPC)}

For solving the above problems, we propose a convenient NHPC method. Here, we demonstrate the new Hadamard `pipeline' coding method of generating Hadamard derived coding, where we use the one-input-four-output coding rules of four fixed quadruple extension transformations such that any derived coding of that Hadamard sequence will be generated directly.

The specific protocol steps of the NHPC method are as follows:

Step 1, create the one-input-four-output coding rules which we called Hadamard pipeline encoder (HPE). And the HPE has four transformation rules of fixed quadruple extension:
\begin{equation} \label{hpe}
 H_{PE}(:,:,i)=\left\{
 \begin{array}{l}
 H_{PE}(:,:,1)=\left[ \begin{array}{cc}
                   +H & +H \\
                   +H & +H
                 \end{array}
                 \right]\\
                 \\
 H_{PE}(:,:,2)=\left[ \begin{array}{cc}
                   +H & +H \\
                   -H & -H
                 \end{array}
                 \right]\\
                 \\
 H_{PE}(:,:,3)=\left[ \begin{array}{cc}
                   +H & -H \\
                   +H & -H
                 \end{array}
                 \right]\\
                 \\
 H_{PE}(:,:,4)=\left[ \begin{array}{cc}
                   +H & -H \\
                   -H & +H
                 \end{array}
                 \right],
 \end{array}
  ~i=1,2,3,4.
  \right.
 \end{equation}

Step 2, we set up a matrix $H_0(1,1,1)$ with a value of $+1$:
\begin{equation}\label{h0}
H_0(1,1,1)=[+1].
\end{equation}

Step 3, bringing Eq.~\ref{h0} into the HPE equation (Eq.~\ref{hpe}), and the results can be expressed as:
\begin{equation} \label{h11}
 H_{PE}(:,:,1)=\left[ \begin{array}{cc}
                   +H_0(1,1,1) & +H_0(1,1,1) \\
                   +H_0(1,1,1) & +H_0(1,1,1)
                 \end{array}
                 \right]=\left[ \begin{array}{cc}
                   +1 & +1 \\
                   +1 & +1
                 \end{array}
                 \right],
 \end{equation}
 \begin{equation}\label{h12}
 H_{PE}(:,:,2)=\left[ \begin{array}{cc}
                   +H_0(1,1,1) & +H_0(1,1,1) \\
                   -H_0(1,1,1) & -H_0(1,1,1)
                 \end{array}
                 \right]=\left[ \begin{array}{cc}
                   +1 & +1 \\
                   -1 & -1
                 \end{array}
                 \right],
  \end{equation}
 \begin{equation}\label{h13}
 H_{PE}(:,:,3)=\left[ \begin{array}{cc}
                   +H_0(1,1,1) & -H_0(1,1,1) \\
                   +H_0(1,1,1) & -H_0(1,1,1)
                 \end{array}
                 \right]=\left[ \begin{array}{cc}
                   +1 & -1 \\
                   +1 & -1
                 \end{array}
                 \right],
  \end{equation}
 \begin{equation}\label{h14}
 H_{PE}(:,:,4)=\left[ \begin{array}{cc}
                   +H_0(1,1,1) & -H_0(1,1,1) \\
                   -H_0(1,1,1) & +H_0(1,1,1)
                 \end{array}
                 \right]=\left[ \begin{array}{cc}
                   +1 & -1 \\
                   -1 & +1
                 \end{array}
                 \right].
\end{equation}

Here, we obtain a martix ($2\times 2~\textrm{pixels}$) with the value of $+1$ and three matrices ($2\times 2~ \textrm{pixels}$) with different $+1$ and $-1$ distributions. In order to obtain the derived codes of subsequent different distributions, we express Eqs.~(\ref{h12})-(\ref{h14}) as $H_1(2,2,j_1),~j_1=1,2,3.$

Step 4, bringing $H_1(2,2,j_1),~j_1=1,2,3.$ into the HPE equation (Eq.~\ref{hpe}). To make  Eq.~\ref{hpe} simple, The HPE equation is abbreviated into $H_{PE}(:,:,i)\{ \},~i=1,2,3,4$.  Hence, we get the 12 kinds of transformation results which can be expressed as:
\begin{equation} \label{hpe1}
 H_{PE}(:,:,i) \{H_1(2,2,j_1)\},~i=1,2,3,4;~j_1=1,2,3.
 \end{equation}

And, Eq.~\ref{hpe1} can be further expressed as:
\begin{equation} \label{pe1}
 H_2(4,4,j_2),~j_2=1,2,\cdots,12~(3\times 4).
 \end{equation}

 Step 5, the resulting $H_2(4,4,j_2),~j_2=1,2,\cdots,12$ is then input into HPE equation Eq.~\ref{hpe}, we also get 48 kinds of transformation results $H_3(8,8,j_3),~j_3=1,2,\cdots,48~(3\times 4^2)$. The output transformation results of the HPE is then input to the HPE for expansion, and repeat this operation until all output results reach the required number. The output results of per HPE can be expressed as:

 \begin{equation}
 H_l(2^l,2^l,j_l),~j_l=1,2,\cdots,3\times 4^{l-1}~(l\geq1),
 \end{equation}
where, $l$ is the number of complete transformations through HPE. And the number of all output results through the HPE is:

 \begin{eqnarray}
 M&=&1+3\times 4^0+3\times 4^1+\cdots+3\times 4^{l-1}\\
 &=&1+\sum^l_{t=1}3\times 4^{t-1}\\
 &=&2^{2\times l},
 \end{eqnarray}

  This is equivalent to the Hadamard derived coding of $2^{2\times l}$ order for THDC.

\section{Results and Discussion}

\begin{figure}
\centering
\includegraphics[width=9.5cm]{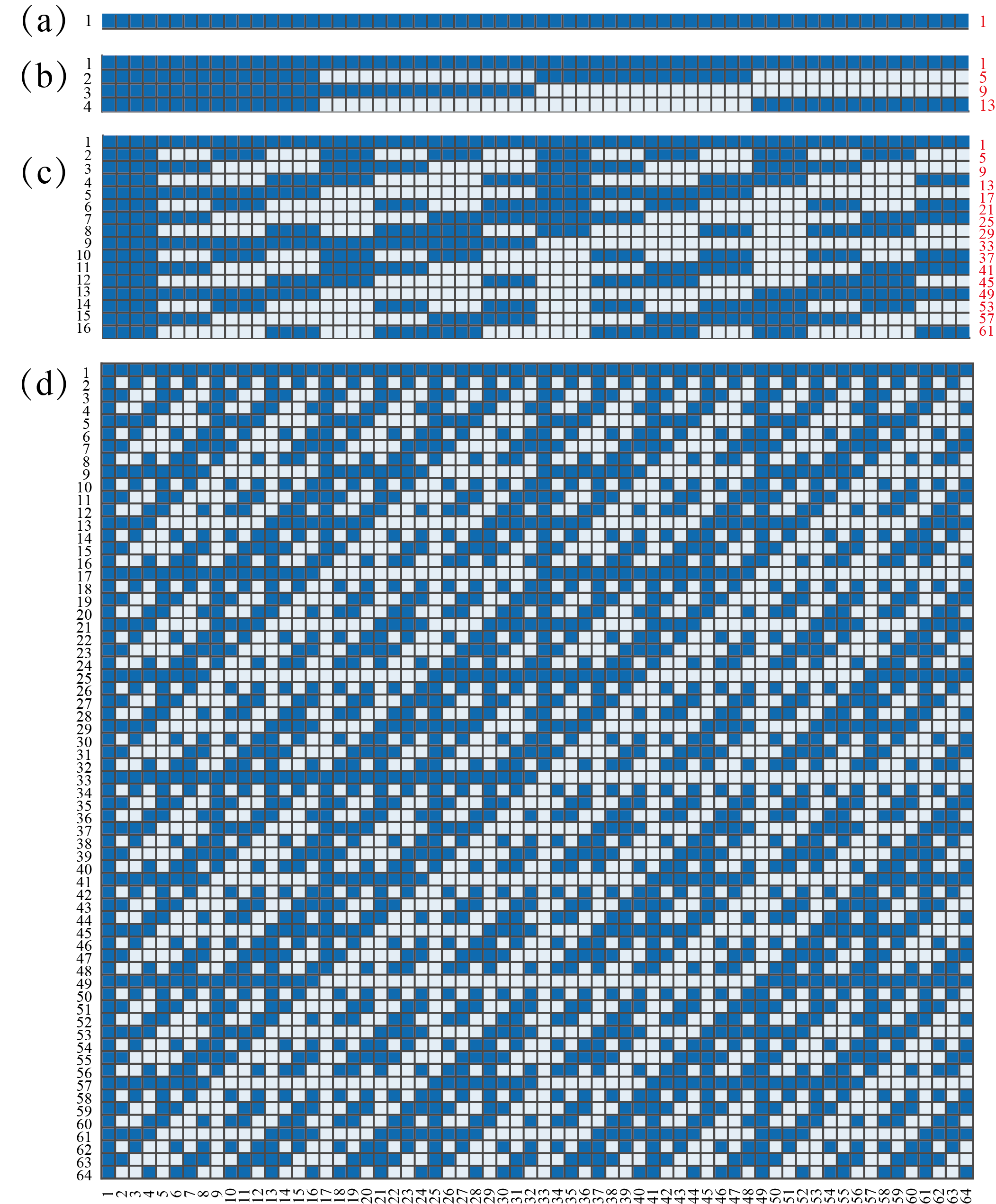}
\caption{The process matrix diagram of THDC method for MPCGI. (a) the Hadamard matrix $H_{1\times 64}$ of size $1\times 64$ pixels; (b) the Hadamard matrix $H_{4\times 64}$ of size $4\times 64$ pixels; (c) the Hadamard matrix $H_{16\times 64}$ of size $16\times 64$ pixels; (d) the Hadamard matrix $H_{64}$ of size $64\times 64$ pixels.}
\label{mpc}
\end{figure}

In order to illustrate the advantages of our NHPC method, we take two classical optimal ordering methods \cite{sun2017russian,Zhou_2019} as examples to compare and analyze.

\subsection{ Multi-resolution progressive computational ghost imaging}
\label{tmpcgi}
\emph{In the traditional THDC method}, the implementation of MPCGI multi-resolution Hadamard sequence needs to be roughly divided into the following steps. Here we take $1\times 1 (2^{2\times 0})$ to $8\times 8 (2^{2\times 3})$ resolution multi-resolution imaging as an example.

Step 1: Generate the Hadamard matrix of order $2^{2\times 3} [H_{64}]$, which is expressed as $H_{64}$ [as shown in Fig.~\ref{mpc}(d)];

Step 2: Generate the Hadamard matrix of order $2^{2\times 2}$ [$H_{16}$]. The direct product of $H_{16}$ and $1\in R^{4\times1}$ is extended to the same number of columns as the $H_{64}$, i.e., we get a matrix $H_{16\times 64}$ of size $16\times 64$ pixels, as shown in Fig.~\ref{mpc}(c). Look for the row vector matrix of $H_{64}$ including $H_{16\times 64}$, and arrange it in front;

Step 3: Generate the Hadamard matrix of order $2^{2\times 1}$ [$H_{4}$]. The direct product of $H_{4}$ and $1\in R^{16\times1}$ is extended to the same number of columns as the $H_{64}$, i.e., we get a matrix $H_{4\times 64}$ of size $4\times 64$ pixels, as shown in Fig.~\ref{mpc}(b). Look for the row vector matrix of $H_{64}$ including $H_{4\times 64}$, and arrange it in front;

Step 4: Generate the Hadamard matrix of order $2^{2\times 0}$ [$H_{1}$]. The direct product of $H_{1}$ and $1\in R^{64\times1}$ is extended to the same number of columns as the $H_{64}$, i.e., we get a matrix $H_{1\times 64}$ of size $1\times 64$ pixels, as shown in Fig.~\ref{mpc}(a). Look for the row vector matrix of $H_{64}$ including $H_{1\times 64}$, and arrange it in front;

Step 5: To construct a modulation matrix of the light field for GI, we will select an arbitrary row of $H_{64}$ to obtain a two-dimensional Hadamard derived pattern.

Step 6: According to the sequence of steps 1 to 5, the Hadamard sequence of MPCGI can be achieved. The Hadamard sequence matrix are expanded to appropriate size by direct product according to actual needs.

\begin{figure}
\centering
\includegraphics[width=10cm]{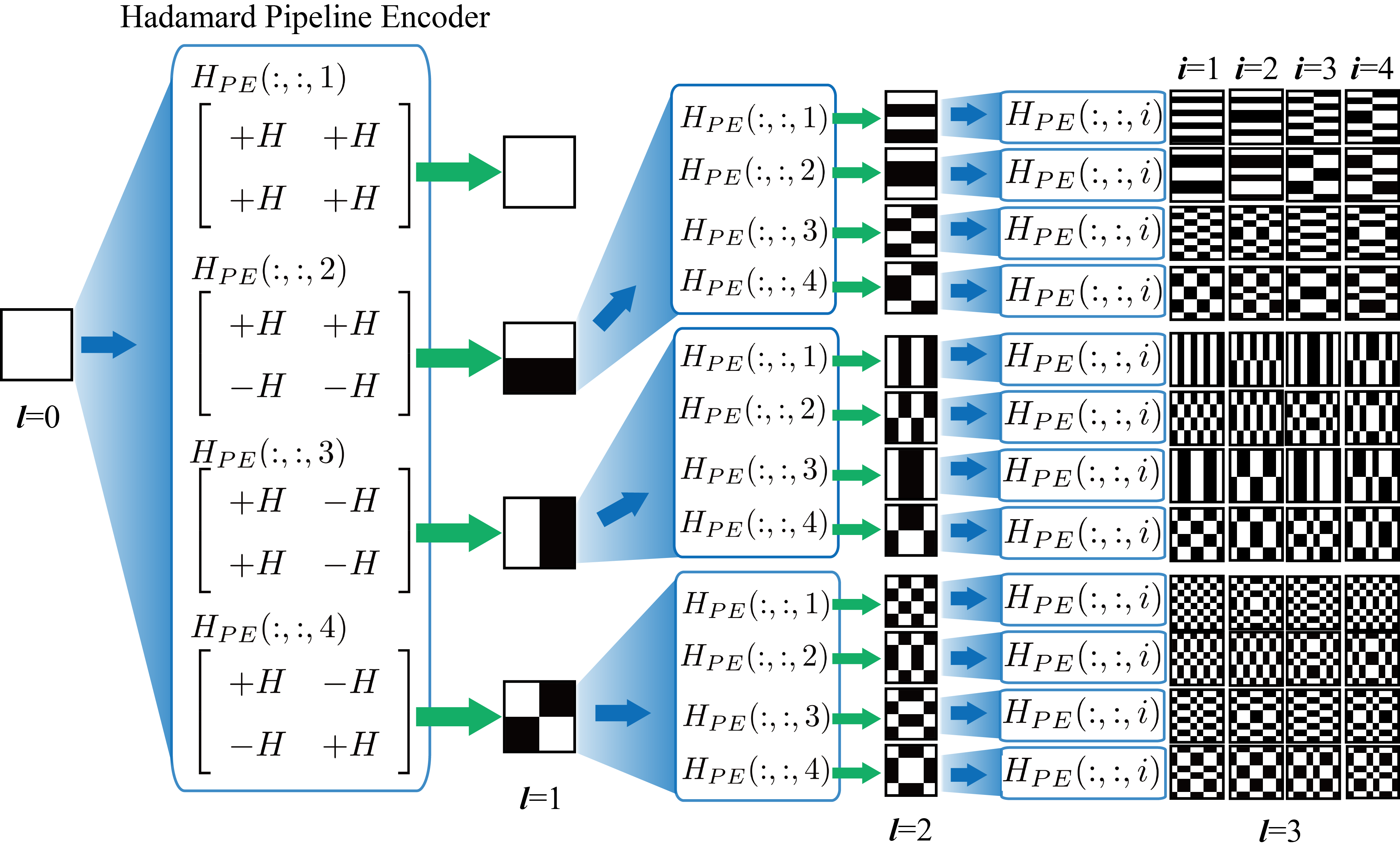}
\caption{The schematic diagram of Hadamard `Pipeline' coding for MPCGI.}
\label{nhpc}
\end{figure}

\emph{In the NHPC method}, the implementation of MPCGI multi-resolution Hadamard sequence needs to be roughly divided into the following steps:

Step 1: Set the initial value and the number of times $l$ to use the HPE. Here, The initial value defaults to 1. According to the NHPC rules, the Hadamard sequence of MPCGI can be obtained by setting $l=3$, i.e., the output image of $l=0$ to $l=3$ per NHPC method (there is no requirement for the sequence of the  four transformation rules inside the HPE), as shown in Fig.~\ref{nhpc}.

Step 2: According to the actual ghost imaging experiment, the Hadamard sequence matrix are extended to the appropriate size by direct product.

\subsection{ A Russian Dolls ordering of the Hadamard basis for compressive single-pixel imaging}

To further illustrate the performance of NHPC method, We chose the Russian Dolls (RD) ordering sequence with more fine arrangement for discussion.

\begin{figure}
\centering
\includegraphics[width=9.5cm]{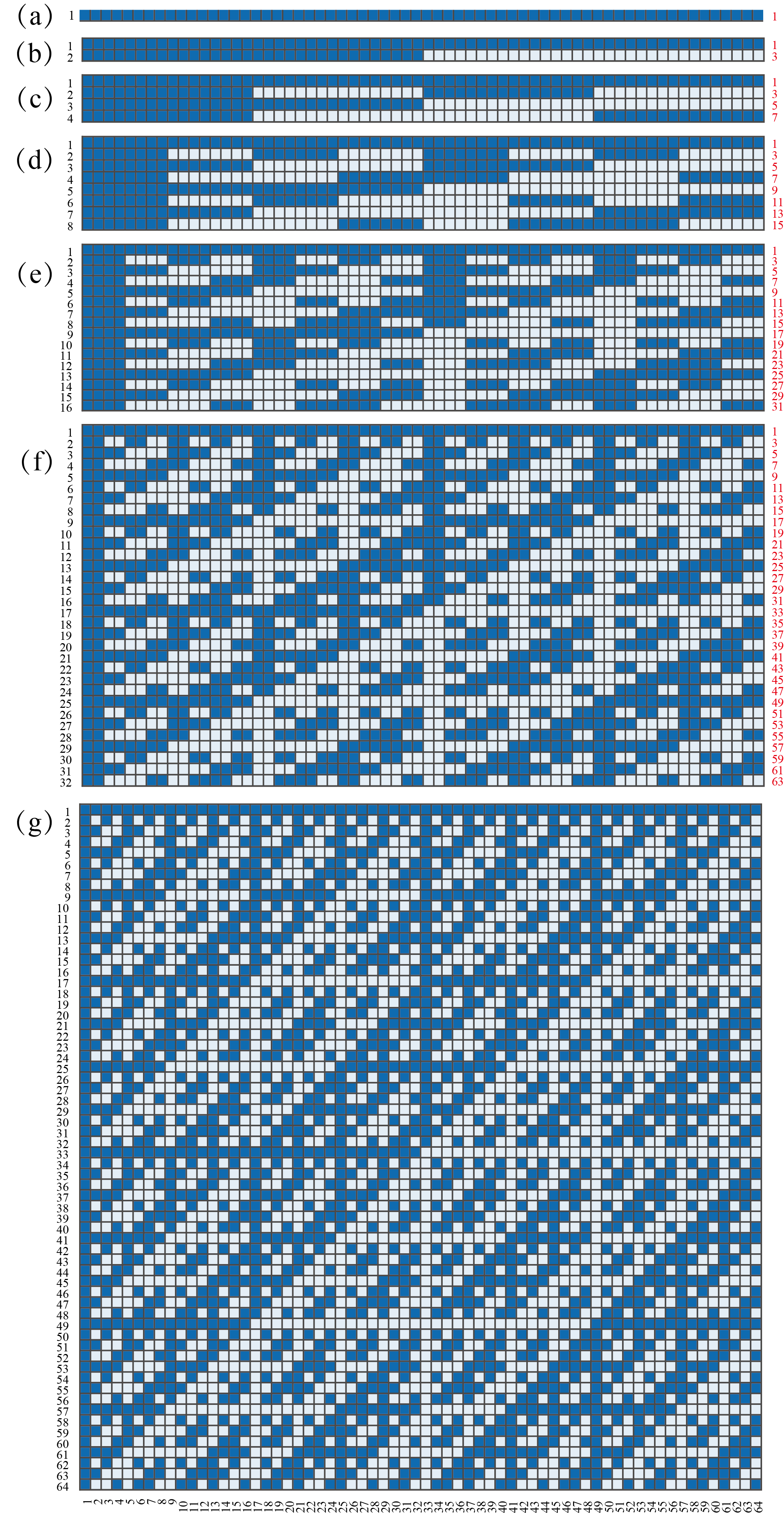}
\caption{(Color online) The process matrix diagram of THDC method for RD ordering. (a) the Hadamard matrix $H_{1\times 64}$ of size $1\times 64$ pixels; (b) the Hadamard matrix $H_{2\times 64}$ of size $1\times 64$ pixels; (c) the Hadamard matrix $H_{4\times 64}$ of size $4\times 64$ pixels; (d) the Hadamard matrix $H_{8\times 64}$ of size $8\times 64$ pixels; (e) the Hadamard matrix $H_{16\times 64}$ of size $16\times 64$ pixels; (f) the Hadamard matrix $H_{32\times 64}$ of size $32\times 64$ pixels; (g) the Hadamard matrix $H_{64}$ of size $64\times 64$ pixels.}
\label{rdscheme}
\end{figure}

Unlike the MPCGI method which sorts $H_{2^{k}}~(k=0,2,4,6,\cdots)$, RD method needs to sort $H_{2^{k}}~(k=0,1,2,3,\cdots)$ from high to low order. Hence, it is necessary to generate all [$H_{2^{k}}~(k=6,5,4,3,2,1,0)$] orders of Hadamard matrix to realize RD ordering sequence by THDC method. Fig.~\ref{rdscheme} is the result of all lower order Hadamard matrices [$H_{2^{k}}~(k=5,4,3,2,1,0)$] being extended by direct product to the same number of columns as order $H_{2^{6}}~[H_{64}]$, which are expressed as $H_{32\times 64}$ [Fig.~\ref{rdscheme}(f)],~$H_{16\times 64}$ [Fig.~\ref{rdscheme}(e)],~$H_{8\times 64}$ [Fig.~\ref{rdscheme}(d)],~$H_{4\times 64}$ [Fig.~\ref{rdscheme}(c)],~$H_{2\times 64}$ [Fig.~\ref{rdscheme}(b)],~$H_{1\times 64}$ [Fig.~\ref{rdscheme}(a)] correspondingly. Then, find the row vector matrix containing lower order matrix $[H_{32\times 64}\sim H_{1\times 64}]$ in $[H_{64}]$ and rank them first according to the order from small to large. Finally, perform the same procedure as in step 5 and step 6 in Sec. \ref{tmpcgi}, and the RD imaging experiment can be implemented. Compared with the MPCGI sequence in Sec. \ref{tmpcgi}, the THDC method to implement RD ordering sequence requires the generation of Hadamard matrix with the order of odd power of 2 ($H_{2^{k}},~n=1,3,5,\cdots$) in the middle of the even power of 2 ($H_{2^{k}},~n=2,4,6,\cdots$), which leads to further increase the generation steps and complexity of implementation.

\begin{figure}
\centering
\includegraphics[width=12cm]{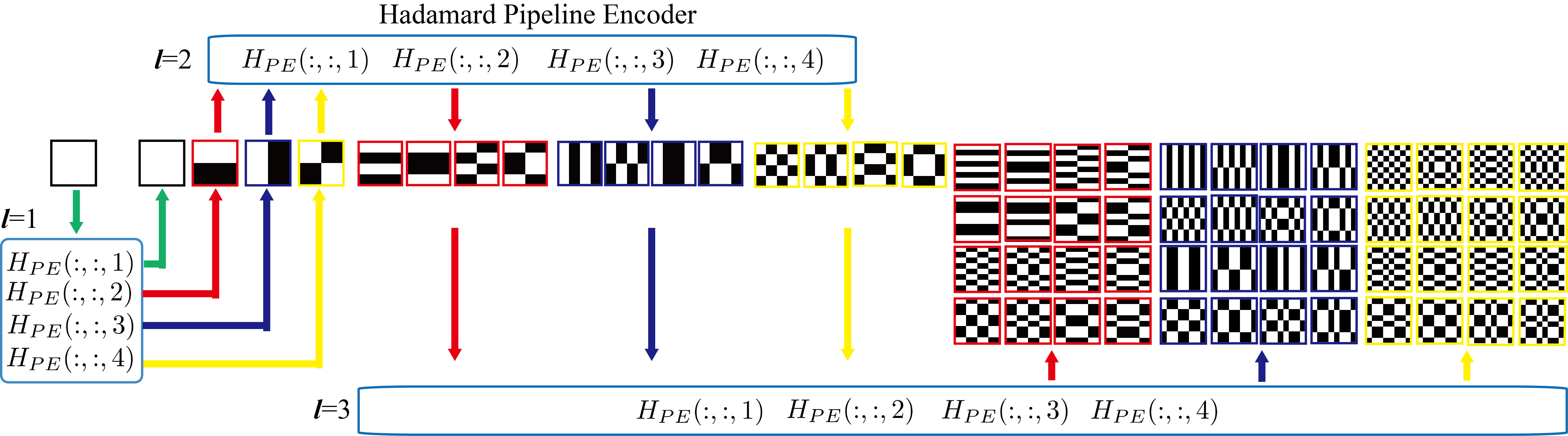}
\caption{The schematic diagram of Hadamard `Pipeline' coding for RD.}
\label{rdgihpe}
\end{figure}

For NHPC method to generate RD ordering sequence, certain restrictions should be added on the basis of MPCGI method, i.e., the sequence of the four transformation rules inside each HPE is fixed [that is $H_{PE}(:,:,1),~
 H_{PE}(:,:,2),~H_{PE}(:,:,3),~H_{PE}(:,:,4)$]. Here, the order of the four transformation rules limited mainly to satisfy the order of odd power of 2 ($H_{2^{k}},~k=1,3,5,\cdots$) and corresponding transpose in RD ordering sequence. Fig.~\ref{rdgihpe} is the schematic diagram of NHPC for RD ordering sequence. The red region of the output image of each HPE is an odd power of 2 ($H_{2^{k}},~k=1,3,5,\cdots$) , and the blue region is its transpose. Therefore, compare with the implementation of MPCGI sequence in Sec. \ref{tmpcgi}, there is only one more restriction, and the generation steps and complexity do not increase. Obviously, this also reflects the advantages of NHPC method.

\subsection{Discussion}

In order to better analyze the performance of NHPC method and its impact on ghost imaging, we compare the memory consumption of runing and the efficiency of generation methods. According to the above work, we also propose a method of directly extracting high-order Hadamard matrix according to index to obtain RD or MPCGI optimal coding.

Here, two methods of generating Hadamard derived optimization matrix are further discussed. We provide the consumption of computer memory of running THDC and NHPC methods with generate the RD ordering sequence, which is represented by the random access memory (RAM) as shown in Table \ref{tabram}. We take the generation of a $H_{2^{k}}~(k=0,1,2,\cdots,K)$ order RD Hadamard sequence as an example. THDC method needs to generate a high-order Hadamard matrix $H_{2^{K}}$, which occupies $2^{K^2}$B of memory. Secondly, all low-order Hadamard matrices $H_{2^{k-1}}$ are generated, which occupies $\sum^{K-1}_{k=1}2^{(k-1)^2}$B of memory. Finally, generate all the extended matrices $H_{2^{(k-1)}\times 2^K}$ corresponding to lower-order matrices $H_{2^{k-1}}$, which occupies $2^{2K-1}$($2^{K}\times \sum^{K-1}_{k=1}2^{(k-1)}$)B of memory. Overall, a total of $2^{K^2}+\sum^{K-1}_{k=1}2^{(k-1)^2}+2^{2K-1}$B memory is required in THDC method. If the Hadamard matrix order is large, it will be difficult to achieve for a computer with limited memory. Our NHPC method performs HPE iteration without the need for intermediate transition matrix to generate RD sequences. As a result, our SGI method requires less than $2^{K^2}$B of memory. In order words, this is a continuous pipeline generation method that does not occupy redundant resources. This method has more advantages in generating Hadamard matrix with higher order. For the above reason, our NHPC method also reduces the hardware requirement for ghost imaging practicality.

\begin{table}[htbp]
\centering
 \caption{\label{tabram}Memory consumption of different methods (RAM/B)}
 \begin{tabular}{lcccc}
  \toprule
       &  $H_{2^{K}}$&  $H_{2^{k-1}}$& $H_{2^{k-1}\times 2^{K}}$& Total \\
  \midrule
 THDC & $2^{K^2}$ & $\sum^{K-1}_{k=1}2^{(k-1)^2}$& $2^{2K-1}$  & $2^{K^2}+\sum^{K-1}_{k=1}2^{(k-1)^2}+2^{2K-1}$ \\
 NHPC & $2^{K^2}$ & 0   & 0 &  $2^{K^2}$\\
  \bottomrule
 \end{tabular}
\end{table}

In order to better make Hadamard matrix used in GI, we further analyzed the optimized Hadamard sequence. First of all, it can be seen from the generation of NHPC method that all two-dimension Hadamrd derived matrices [as shown in Fig.~\ref{nhpc} and Fig.~\ref{rdgihpe}] are generated by 1, HPE and corresponding results through HPE. Similarly, Hadamard matrix also can be generated by 1, $H=\left[+1 , +1 ; +1 , -1\right]$ and corresponding results through $H$, and can be expressed as:
\begin{equation} \label{syh}
   H_{2^k}=\left[1\right]\bigotimes H^{(k=1)}\bigotimes H^{(k=2)}\bigotimes \cdots \bigotimes H^{(k=K)},~k=1,2,\cdots,K.
\end{equation}
Hence, from Eq.~\ref{syh}, it can be seen that all Hadamard matrices have the same distribution of odd rows with the next higher-order matrix, while the even behavior of the higher-order matrix is the new matrix generated by the lower-order Hadamard matrix. For example, the red numbers on the right of Figs.~\ref{rdscheme}(a)-(f) represent the same index row number as the row distribution of the next higher-order matrix, respectively. To be brief, the odd rows of high-order Hadamard matrices (row number:$1,3,5,7,9,11,13,15,\cdots$) is its own first lower-order Hadamard matrix. For an even order Hadamard matrix, the odd behavior of an odd row of a higher-order Hadamard matrix (row number:$1,5,9,13,17,\cdots$) is its own first lower even-order Hadamard matrix, as shown in Figs.~\ref{mpc}(a)-(c). On the basis of this part of work, we can also generate only a high-order Hadamard matrix, extract the odd number of its rows, and arrange them in front to realize RD, MPCGI or other optimized coding.

\section{Conclusion}
In this paper, we present a NHPC method based on four fixed transformation rules to directly generate two-dimensional Hadamard derived pattern for GI. This NHPC method is a simple and convenient light field coding method without redundant operation and consumption. Compared with the commonly used THDC method in GI, both results and discussion have been used to demonstrate its exceptional features. First, the NHPC method requires less memory consumption [reduce $\left(\sum^{K-1}_{k=1}2^{(k-1)^2}+2^{2K-1}\right)$B memory consumption] , which reduces the hardware requirements in GI experiments. Second, this is an efficient and continuous generating optical field coding method which uses four fixed transformation rules and loop iteration to realize the generate subsequent Hadmard derived diagrams as needed directly. Third, according to the properties of Hadamard matrix, the HPE rules can be formulated to directly generate the optimized Hadamard sequence. Therefore, the optimized coding method is simplified, which is more conducive to the design and implementation of Hadamard optimized coding. Furthermore, the NHPC method is also beneficial to hardware realization and further promotes the practical application of ghost imaging.

\section{Funding}

This work is supported by the Project of the Science and Technology Department of Jilin Province (Grant No. 20170204023GX); the Special Funds for Provincial Industrial Innovation in Jilin Province (Grant No. 2018C040-4, 2019C025); the Young Foundation of Science and Technology Department of Jilin Province (20170520109JH); the Science Foundation of the Education Department of Jilin Province ( 2016286, 2019LY508L35).


\end{document}